# Tetragonal Bismuth Bilayer: A Stable and Robust Quantum Spin Hall Insulator

Liangzhi Kou[1, 2, *], Xin Tan[2], Yandong Ma[3], Hassan Tahini[2], Liujiang Zhou[4], Ziqi Sun[1], Aijun Du[1], Changfeng Chen[5], Sean C Smith[2]

[1]School of Chemistry, Physics and Mechanical Engineering Faculty, Queensland University of Technology, Garden Point Campus, QLD 4001, Brisbane, Australia

[2]Integrated Materials Design Centre (IMDC), School of Chemical Engineering, University of New South Wales, Sydney, NSW 2052, Australia

[3]Engineering and Science, Jacobs University Bremen, Campus Ring 1, 28759 Bremen, Germany

[4]Bremen Center for Computational Materials Science, University of Bremen, Am Falturm 1, 28359 Bremen, Germany

[5]Department of Physics and Astronomy and High Pressure Science and Engineering Center, University of Nevada, Las Vegas, Nevada 89154, United States

**Abstract**: Topological insulators (TIs) exhibit novel physics with great promise for new devices, but considerable challenges remain to identify TIs with high structural stability and large nontrivial band gap suitable for practical applications. Here we predict by first-principles calculations a two-dimensional (2D) TI, also known as a quantum spin Hall (QSH) insulator, in a tetragonal bismuth bilayer (TB-Bi) structure that is dynamically and thermally stable based on phonon calculations and finite-temperature molecular dynamics simulations. Density functional theory and tight-binding calculations reveal a band inversion among the Bi-$p$ orbits driven by the strong intrinsic spin-orbit coupling, producing a large nontrivial band gap, which can be effectively tuned by moderate strains. The helical gapless edge states exhibit a linear dispersion with a high Fermi velocity comparable to that of graphene, and the QSH phase remains robust on a NaCl substrate. These remarkable properties place TB-Bi among the most promising 2D TIs for high-speed spintronic devices, and the present results provide insights into the intriguing QSH phenomenon in this new Bi structure and offer guidance for its implementation in potential applications.

Email: Liangzhi.kou@qut.edu.au

**Introduction**

The discovery of topological insulators (TIs) represents one of the most important advances in physics and materials science in recent years, they distinguish from normal insulator or semiconductors by insulating bulk but gapless surface or edge states [1-3]. The protected metallic surface or edge states is robust to the defect and non-magnetic doping, acts as a dissipationless transport channel that are promising for applications in spintronics and quantum computations [4]. Although numerous three-dimensional (3D) TIs have been experimentally and theoretically demonstrated, 2D TIs are rare until now. The first 2D TI was proposed in graphene from analytic models[3]; however, its nontrivial band gap opened by the weak intrinsic spin-orbit coupling (SOC) is too small (~$10^{-3}$ meV) [5], to practically use under normal temperature. Several proposals have been put forward to increase the SOC effect (thus nontrivial gap) [6-10]and new material forms have been theoretically predicted to be QSH insulators[11-12], such as silicene, gemanene and stanene, but only a small number of proposed quantum wells[13-14]have been verified to exhibit TI characters. Extremely low temperature conditions imposed by the small band gaps still impede practical applications, which has stimulated intensive ongoing search for large-gap QSH insulators for realistic implementations and operations of this class of novel materials

Among the theoretically predicted or experimentally confirmed TIs in 2D or 3D, the compounds containing the element bismuth (Bi) are among the most promising due to the strong intrinsic spin-orbit coupling in Bi, which drives electronic band inversions leading to topological insulating phases. Examples include 3D TIs $Bi_2Se_3$ ($Bi_2Te_3$), [15] $BaBiO_3$, [16] and BiTeI, [17] and 2D TIs Bi(111) bilayer, [18-20]bilayers of group III elements with Bi, [21] chemically modified Bi honeycomb lattices, [22-23]as well as $Bi_4Br_4$ thin films. [24] In particular, the 1D topological edge states of Bi bilayer have been observed in scanning tunneling microscopy experiments, providing a direct spectroscopic evidence of the TI nature[25] and confirming the theoretical predictions. [18-20] The Bi bilayer has a strong out-of-plane buckling due to the π-π bonding, and this structural feature is shared in silicene, germanene and stanene[11-12], all of which possess the hexagonal honeycomb structure as in graphene. Non-hexagonal carbon phases have been proposed, e.g., graphenylene, graphyne, and the square lattice[26-28]; they exhibit metallic or semiconducting nature depending on their structural details, which can effectively tune the electronic properties. Another prominent example is the transition-metal dichalcogenides, which are large-gap semiconductors in the hexagonal lattice (2H phase)[29], but turn into TIs in the square-lattice structures[30-32] and twisted structures (1T' phase). [33] We adopt this strategy of electronic modulation by structure in search of new 2D TIs with a robust large nontrivial band gap required for practical applications in nanoscale devices.

In the present work, we identify a new 2D TI state in a tetragonal Bi bilayer (TB-Bi) configuration containing Bi octagons and tetragons with an out-of-plane buckling pattern. The structural stability of TB-Bi is confirmed by phonon calculations and finite-temperature molecule dynamics simulations, and the topological phases are verified by first-principles electronic band calculations based on density functional theory (DFT), which show a band inversion between the $p_x/p_y$ and $p_z$ orbits and explicit gapless helical edge states. A tight-binding model was constructed to corroborate the predictions of the DFT calculations and unveil the underlying mechanisms. The nontrivial band gap in the QSH state of TB-Bi is large enough for room temperature applications, and it is tunable by moderate strains. Moreover, we identify NaCl as a suitable substrate that couple to TB-Bi via the weak van der Waals interactions with little effect on the electronic structure, especially the topological nature, of TB-Bi. It is expected that the same physical phenomena also can be observed in other similar layered materials, such as silicene, germanene and stanene, in the tetragonal lattices. These findings greatly enrich the 2D TI families and provide important guidance for the search and synthesis of additional 2D layered TIs.

**Methods of Calculations**

First-principles calculations based on the density functional theory (DFT) were carried out using the Vienna Ab Initio Simulation Package (VASP). [34] The exchange correlation interaction was treated within the generalized gradient approximation (GGA) in the form proposed by Perdew, Burke, and Ernzerhof (PBE). [35] The atomic positions were relaxed until the maximum force on each atom was less than 0.01 eV/Å. The energy cutoff of the plane waves was set to 400 eV with an energy precision of $10^{-5}$ eV. For the 2D structures, the Brillouin zone (BZ) was sampled by using a 10×10×1 Gamma-centered Monkhorst-Pack grid, while a 10×1×1 grid was used for the nanoribbon. The vacuum space was set to at least 10 Å in the calculations to minimize artificial interactions between the neighboring slabs. The SOC was included by a second variational procedure on a fully self-consistent basis. The phonon frequencies are calculated by using DFPT method [36] as implemented in the PHONOPY code[37].

**Results and Discussions**

We show in Fig 1a the fully relaxed structure of TB-Bi, which has an equilibrium lattice constant $a$=8.743 Å determined from the energy minimization procedure (Fig 1b) In contrast to the commonly studied hexagonal bilayer bismuth (HB-Bi), the new structure comprises Bi tetragons and octagons, and the bond lengths of the nearest-neighbor Bi-Bi at the tetragons and octagons are slightly different, which are 3.059 and 3.044 Å, respectively  As observed in tetragonal graphene [28] and silicene[12], TB-Bi has a buckled structure, i.e., an out-of-plane deformation, along the thickness direction caused by the unsaturated $p_z$ orbit  The buckling height of 1.76 Å is significantly larger than the corresponding value (1.44 Å) in HB-Bi To confirm the dynamical and thermal stability of TB-Bi, we have performed phonon dispersion calculations and run finite-temperature molecule dynamic simulations The results in Fig 1c show that all the phonon branches have positive frequencies with no imaginary phonon modes in the free-standing TB-Bi, thus indicating the dynamical stability of the structure For the ab initio molecular dynamics simulations, we adopted a relatively large supercell of 3×3 unit cells with the dimensions of 25.63 Å×25.63 Å Structural snapshots of the TB-Bi supercell at the end of the simulation runs (5 ps) are shown in Figs 1d and 1e The results indicate that the TB-Bi structure remains intact at 300 K and 400 K, although wrinkles and distortions develop in the thickness direction It is noted that the formation energy for TB-Bi is -4.2485 eV/atom, which is slightly less than the value (-4.4548 eV/atom) for the hexagonal Bi bilayer However, the dynamical and thermal stability of TB-Bi establishes the viability of this structure and suggests the feasibility of its experimental synthesis

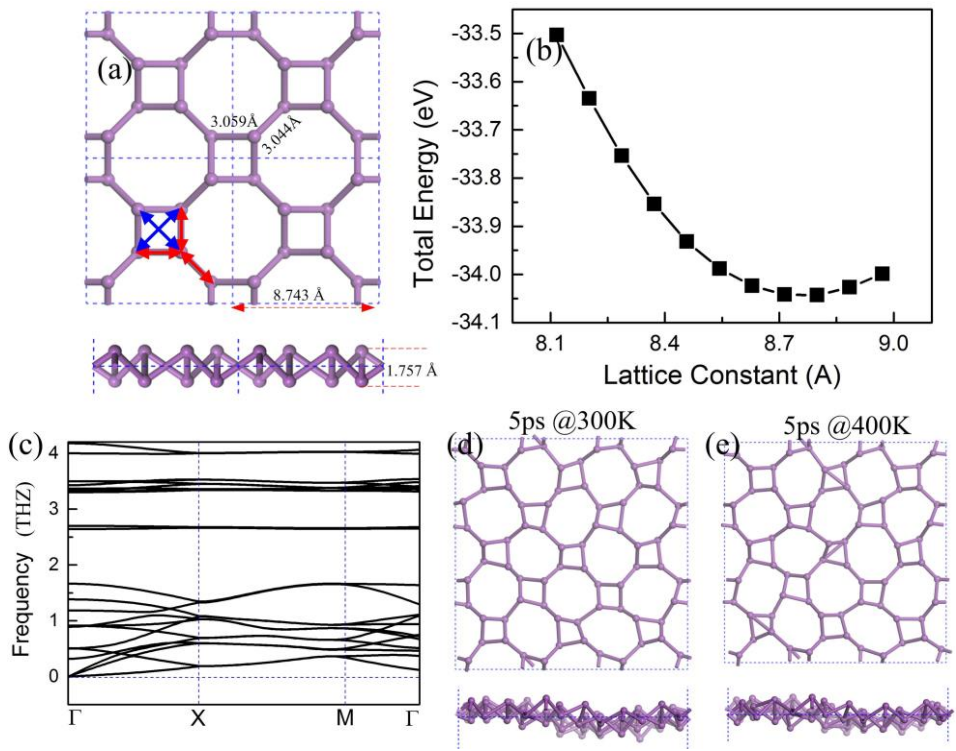

Figure 1 (a) Top and side view of the TB-Bi structure The blue dashed square indicates the unit cell of TB-Bi, the solid red arrows indicate the nearest-neighbor interactions between the Bi atoms in octagons and tetragons, while the solid blue arrows indicate the second nearest-neighbor interactions between the Bi atoms in tetragons (b) Total energy vs the lattice constant for TB-Bi (c) Calculated phonon dispersion for TB-Bi (d,e) Structural snapshots of a TB-Bi supercell from the molecule dynamics simulations at 300 K and 400 K.

We now turn to the electronic properties of TB-Bi. We first calculated the electronic band structure without considering the SOC effect. The obtained results shown in Fig 2(a) indicate that TB-Bi is a semiconductor with a direct band gap of 0.58 eV with both the valence band maximum (VBM) and conduction band minimum (CBM) located at the Γ point. An orbit-projected analysis for the composition of the electronic states in the vicinity of the Fermi level reveals that the states near the VBM arise mainly from the $p_x$ and $p_y$ atomic orbits, while the states near the CBM are largely contributed by the $p_z$ orbits. Such a band alignment has also been reported for the hexagonal Bi bilayer structure. We then turned on the SOC to assess its effect on the electronic band structure. The calculated results shown in Fig 2b reveal significant changes compared to the non-SOC band structure. The most pronounced is the development of a Mexican-hat shape near the top of the VBM, which is caused by an SOC-driven band inversion (see below). A band gap of 0.41 eV (the global indirect gap is 0.3 eV) is opened up at the Γ point by the very strong SOC effect in Bi (about 1 eV). This SOC-driven gap is large enough for realizing the elusive QSH state and implementing device applications at room temperature. The band inversion is clearly visible in the orbit-resolved band structures. With the SOC in consideration (Fig 2b), the CBM states are mainly contributed by the $p_x$ and $p_y$ orbits and the VBM primarily comprises states from the $p_z$ orbits, which are opposite to the band-state alignment in the non-SOC band structures (Fig 2a). Such a band inversion is characteristic of a topological phase transition, which we will further evaluate and confirm below. We have summarized the inversion and the parity exchange in the inset of Fig. 2, where the atomic energy eigenvalue evolution at the Γ point under the effect of the crystal-field splitting and SOC is presented. As the states near the Fermi surface are mainly coming from the p orbitals, we neglect the effect of the s orbitals. Due to the chemical bonding between Bi atoms, the p orbits are split in the first stage. In the second stage, we consider the effect of the crystal-field splitting between different p orbitals. According to the point-group symmetry, the pz orbital is split from the px and py orbitals whereas the last two remain degenerate. When the SOC effect is taken into the account, it drives the band inversions between the pz and the $p_{x,y}$ states.

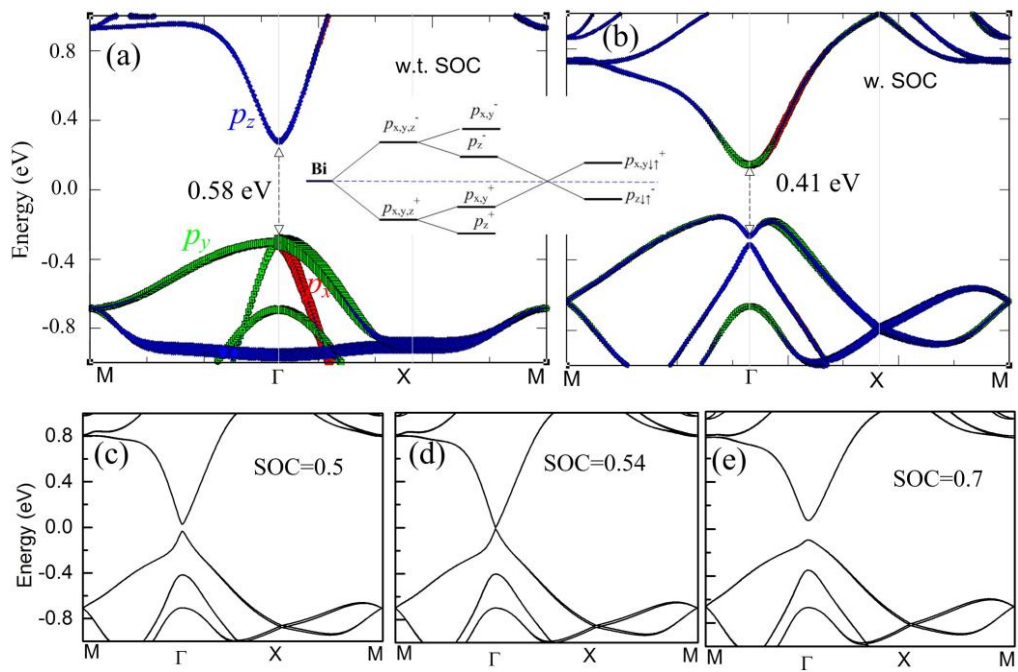

Figure 2 Top panels: Calculated electronic band structures of TB-Bi (a) without and (b) with the spin-orbit coupling (SOC) effect. A schematic diagram illustrating the evolution from the atomic px,y,z orbitals of Bi into the conduction and valence bands at the Γ point is shown in the inset. The red, green and blue symbols represent the contributions from the $p_x$, $p_y$ and $p_z$ orbits of bismuth Bottom panels: The electronic band structures from the tight-binding calculations with the SOC parameter $\lambda_{SOC}$ of (c) 0.00 eV, (d) 0.368 eV, and (e) 0.49 eV

For a systematic evaluation of the band inversion in TB-Bi, we have artificially tuned the SOC strength and monitored the corresponding change of the calculated band gap, which is an effective approach for assessing phase transitions toward a topological insulating state.[21] We show in Fig 3a the band gap variation versus the relative SOC coupling, which is defined as the ratio of the tuned SOC to the full SOC strength in the DFT calculations. It is seen that increasing SOC supresses the band gap, leading to a complete band-gap closure, and then reopens the band gap as the SOC further increases and approaches the full realistic value in TB-Bi. The calculated band structures at three representative tuned SOC values are presented in Fig 3c-3e, where the band-gap reduction, closure, and reopening are illustrated. This band-gap closure-reopening process occurs concomitantly with a band-state character switch between the CBM and VBM as demonstrated by the orbit-resolved band structure shown in Fig 2. These results confirm that TB-Bi is a 2D topological insulator with a nontrivial DFT gap of 0.41 eV, which is larger than the corresponding values of the prominent 3D TI $Bi_2Se_3$ (0.3 eV) and most of the existing 2D TIs, such as silicene, germanene, and stanene.

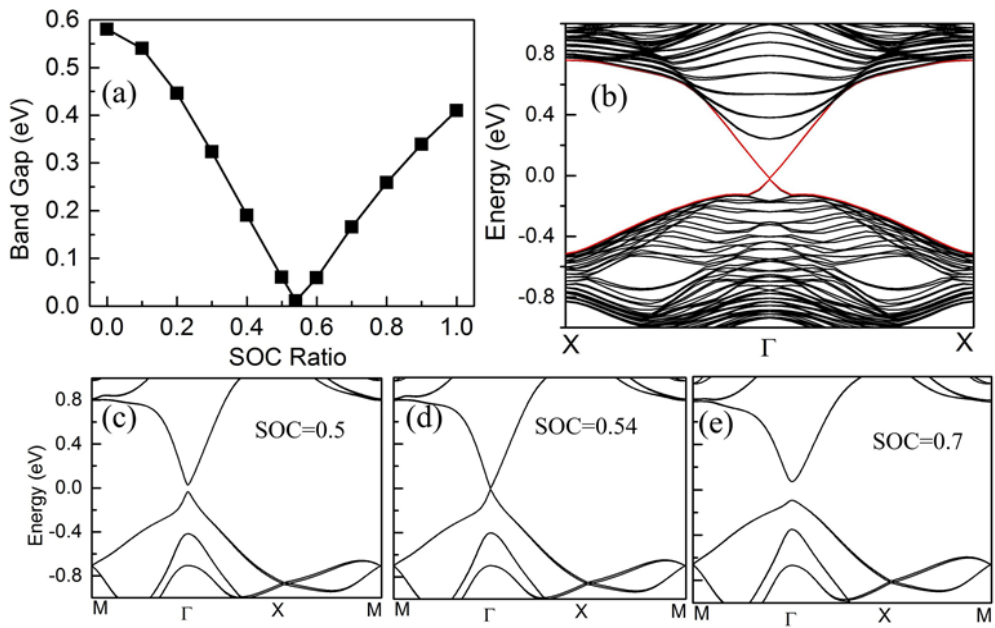

Figure 3 (a) The DFT band gap of TB-Bi as a function of the artificially tuned SOC ratio, which shows an SOC-driven band-gap closure and reopening, indicating a band inversion and transition to the topological insulating phase (b) A demonstration of the edge states in a TB-Bi nanoribbon (c-e) The calculated electronic band structures of TB-Bi at three representative relative SOC coupling of 0.5, 0.54 and 0.7, respectively.

It is well known that standard DFT calculations often underestimate the band gap of semiconductors or insulators. For an improved estimate of the band gap, we have performed additional calculations using the hybrid HSE06 functional, which has produced better agreements with experimental results for many materials. The calculations show that the non-SOC band gap is increased 0.91 eV, and the full-SOC band gap is now 0.22 eV, which is smaller than the value (0.41 eV) predicted by the PBE-GGA calculations, but it is still large enough for room-temperature applications. Most important, the nontrivial band topology is still preserved by the large SOC strength of Bi, see supporting information (the band inversion occurs at SOC=0.7).

To further assess the band inversion mechanism between the Bi-*p* orbits, we constructed a tight-binding model including the $p_x$, $p_y$, and $p_z$ orbits of the Bi atoms to fit the DFT results. The effective Hamiltonian without the SOC effect reads[38]:

$$H = \sum_{i,\alpha} \varepsilon_i^\alpha c_i^{\alpha\dagger} c_i^\alpha + \sum_{<i,j>,\alpha,\beta} t_{ij}^{\alpha\beta} (c_i^{\alpha\dagger} c_j^\beta + h.c.)$$

Where $\varepsilon_i^\alpha$, $c_i^{\alpha\dagger}$, and $c_i^\alpha$ represent the on-site energy, creation, and annihilation operators of an electron at the *α*-orbit of the *i*-th Bi atom Here we consider the nearest-neighbour and the next-nearest-neighbour hopping terms, which can be evaluated by the following expressions:

$$t_{ij}^{p_x p_x} = V_{pp\sigma} \times \cos^2\theta + V_{pp\pi} \times \sin^2\theta$$

$$t_{ij}^{p_x p_y} = (V_{pp\sigma} - V_{pp\pi}) \times \cos\theta \times \cos\varphi$$

$$t_{ij}^{p_x p_z} = (V_{pp\sigma} - V_{pp\pi}) \times \cos\theta \times \cos\gamma$$

where *θ*, *φ* and *γ* are the angles of the vector pointing from the *i*-th atom to the *j*-th atom with respect to the x, y and z axis, respectively. For the nearest-neighbour hopping, the *p*-orbit interaction parameters are set to $V_{pp\sigma}$ = 1.8 eV and $V_{pp\pi}$ = −0.6 eV for the σ and π orbits, and the next-nearest-neighbour hopping parameters are scaled by a factor 0.45. [39-41]The non-SOC band structure is shown in Fig 2c, which is in good agreement with the DFT results, especially the direct band gap at the Γ point. This shows that the low-energy bands near the Fermi level is well described by the Bi-p orbits, while the s-orbit is located deep in the energy spectrum with little influence on the electronic properties near the Fermi level.

To include the SOC effect in our tight-binding calculations, we consider an on-site term:

$$H^{SOC}_{i,\alpha\beta} = \lambda_{SOC}\langle \vec{L}\ \vec{\sigma}\rangle_{\alpha\beta}$$

where $\vec{L}$ is the angular momentum operator and $\vec{\sigma}$ is the Pauli matrix. The matrix element is evaluated in the basis of the atomic orbits ($p_x$, $p_y$, $p_z$) of the $i$-th atom, and $\lambda_{SOC}$ is the SOC strength of the Bi atom. Our calculations show that increasing SOC strength $\lambda_{SOC}$ initially reduces the band gap at the Γ point, which is consistent with the results of the DFT calculations. When $\lambda_{SOC}$ reaches 0.368 eV, the valence and conduction bands meet at a single point at the Γ point (Fig 2d), resulting in a complete band-gap closure. Further increased $\lambda_{SOC}$ reopens the band gap and eventually produces a band structure similar to the DFT results (Fig 2b) at $\lambda_{SOC}$ = 0.49 eV (Fig 2e). It is noted that only the on-site SOC effect is considered in the present tight-binding calculations since it is the dominant term [23, 42-43]

A key characteristic feature of a 2D TI is the presence of helical gapless edge states that offer a novel transport channel for low-dissipation carrier conduction. To demonstrate such QSH states in TB-Bi, we examine an armchair-edged TB-Bi nanoribbon as a representative case study without any loss of universality. The edge Bi atoms are hydrogenated to eliminate the dangling bonds, and the width of the nanoribbon is set to a large value of 8.6 nm to avoid interactions between the edge states on the opposite sides. The calculated band structure of the nanoribbon is shown in Fig 3b, where the gapless edge states appear inside the bulk gap and cross linearly at the Γ point. The Fermi velocity of these edge states is estimated to be

$6.34\times10^5$ m.s$^{-1}$, which is closer to the value $8.46\times10^5$ m.s$^{-1}$ in graphene [12], which renders TB-Bi promising used in high-speed spintronic devices.

Strain engineering is a powerful approach to modulating the electronic properties and the topological nature of 2D materials, as demonstrated in the topological phase transition and gap modulation in the Sb(111) bilayer induced by a biaxial lattice expansion[44]. Here we take this approach to investigate the effect of a biaxial strain on the topological properties of TB-Bi. We impose a biaxial strain by tuning the planar lattice parameter, and then re-optimize the atomic positions at each set lattice parameter. The magnitude of the strain is described by $\varepsilon = \Delta a/a_0$, where $a_0$ and $a = \Delta a + a_0$ denote the lattice parameters of the unstrained and strained systems, respectively. Our calculations show that the nontrivial band topology is preserved up to at least ±5% strain. Such a robust topological nature of the electronic band structure against lattice deformation makes it easier for experimental realization and characterization of TB-Bi as a new 2D TI with QSH states. Meanwhile, for potential applications, it is also of considerable interest to effectively tune the topological properties by external control. We show in Fig 4a the magnitude of nontrivial the global bulk band gap as a function of strain It is seen that the band gap decreases with the increasing tensile strain, closing at 5% strain, while a compressive strain slightly enhances the gap.

Substrate effect is an important consideration in nanomaterial synthesis and device fabrication, and it may adversely impact the properties of 2D Tis [45]. Previous investigations demonstrated that silicene can be turn from a 2D TI into a trivial semiconductor due to its strong active chemical surface [40]. A hexagonal Bi lattice grown on the Si(111) surface functionalized with one-third monolayer halogen atoms exhibits an isolated QSH state with an energy gap as large as 0.8 eV resulting from a substrate-orbital-filtering effect [46]. To explore the substrate effect on the topological phase of TB-Bi, we considered a superstructure of 2×2 TB-Bi on a crystalline NaCl substrate (modelled by a two-layer 3×3 supercell with the bottom layer fixed in the bulk atomic position), as shown in Fig 4b. The lattice mismatch between TB-Bi and the NaCl substrate is about 2.8%, which will not alter the band topology as shown above. The optimized interlayer space between TB-Bi and the substrate is 3.46 Å, which is in the range of weak van der Waals interactions. Calculations show that the electronic bands of the freestanding TB-Bi near the Fermi level are well preserved in the superstructure, as shown in Fig 4c. The SOC band gaps are barely affected by the NaCl substrate, which implies that NaCl may serve as a suitable template for TB-Bi in the process of chemical growth and device fabrications. With the synthesized TB-Bi bulk material growth on NaCl crystal, we built a model containing 5 bilayers of tetragonal Bi to quantify the binding energy, and the calculations indicate that the energy of removing the first surface bilayer is 0.389 eV, which is 85.3 mJ/m². This exfoliation energy is comparable to the values for graphene and transition metal dichalcogenides (65-120 mJ/m2) [47], thus it is feasible to exfoliate the tetragonal Bi bilayer from its bulk phase.

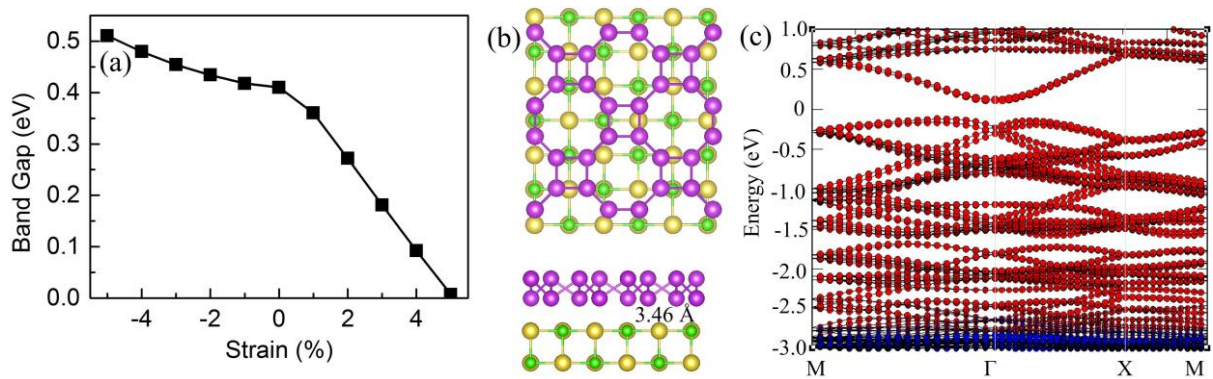

Figure 4 (a) The modulation of the nontrivial band gap by applied strains within the range of -5% to 5% (b) The structural models of TB-Bi on top of a NaCl substrate (c) The band structures of TB-Bi@NaCl, where the red dotted lines indicate states from bismuth while blue lines are those from the NaCl crystal

**Conclusions**

We have predicted by first-principles calculations a robust QSH phase in a tetragonal bilayer Bi structure, termed TB-Bi, which is stable beyond the room temperature. An electronic band inversion between the $p_x/p_y$ and $p_z$ orbits is driven by the strong intrinsic SOC in Bi, producing in TB-Bi a topologically nontrivial state with a band gap that is suitable for room-temperature device applications. We also performed tight-binding calculations that offer insights into the underlying mechanisms responsible for the band inversion. The helical edge states in TB-Bi possess a very high Fermi velocity comparable to that of graphene, which bodes well for its applications in high-speed spintronic devices. The present results call for experimental efforts in synthesis and characterization of the TB-Bi structure, which may stimulate a broad range of activities in new 2D TI research and development.

**Acknowledgement**

Financial support by the ARC Discovery Early Career Researcher Award (DE150101854) is gratefully acknowledged. The useful discussion and help to build tight binding model from Dr. Linyang Li are acknowledged. This research was undertaken with assistances provided at the NCI National Facility systems at Australia National University through the National Computational Merit Allocation Scheme Supported by the Australia Government. C.F.C was partially supported by the US Department of Energy through the Cooperative Agreement DE-NA0001982.

**References**

[1]  Hasan M Z, Kane C L, 2010, *Rev Mod Phys* **82**, 3045.

[2]  Qi X L, and Zhang S C, 2011 *Rev Mod Phys*., **83**, 1057.

[3]  Kane C L, and Mele E J, 2005 *Phys Rev Lett*., **95**, 226801.

[4]  Bernevig B A, Hughes T L and Zhang S C, 2006 *Science*, **314**, 1757.

[5]  Yao Y, Ye F, Qi X L, Zhang S C, Fang Z, 2007 *Phys Rev B,* **75**, 041401(R).

[6]  Weeks C, Hu J, Alicea J, Franz M, Wu R, 2001 *Phys Rev X*, **1**, 021001.

[7]  Kou L, Hu F, Yan B, Wehling T, Felser C, Frauenheim T, and Chen C, 2015 *Carbon*, **87**, 418-423.

[8]  Kou L, Wu S C, Felser C, Frauenheim T, Chen C, and Yan B, 2015 *ACS Nano*, **8**, 10448-10454.

[9]   Kou L, Yan B, Hu F, Wu S C, Wehling T, Felser C, Chen C, and Frauenheim T, 2013 *Nano Lett*, **13**, 6251-6255.

[10]  Kaloni T P, Kou L, Frauenheim T, Schwingenschlögl U, 2014 *Appl Phys Lett* **105**, 233112.

[11]  Xu Y, Yan B, Zhang H J, Wang J, Xu G, Tang P, Duan W, and Zhang S C, 2013 *Phys Rev Lett* **111**, 136804.

[12]  Liu C C, Feng W, and Yao Y, 2011 *Phys Rev Lett* **107**, 076802.

[13]  Konig M, Wiedmann S, Brune C, Roth A, Buhmann H, Molenkamp L W, Qi X L, and S C Zhang, 2007 *Science*, **318**, 766-770.

[14]  Knez I, Du R R, and Sullivan G, 2011 *Phys Rev Lett*., **107**, 136603.

[15]  Zhang H J, Liu C X,  Qi X L,  Dai X,  Fang Z, Zhang S C, 2009 *Nat Phys* **5**, 438-442.

[16]  Yan B,  Jansen M,  Felser C, 2013 *Nat Phys* **9**, 709-711.

[17]  Bahramy M, Yang B J, Arita R, Nagaosa N, 2012 *Nat Commun* **3**, 679.

[18]  Chen L, Wang Z F, and Liu F, 2013 *Phys Rev B*, **87**, 235420.

[19]  Yang F, Miao L, Wang Z F, Yao M Y, Zhu F, Song Y R, Wang M X, Xu J P, Fedorov A V, Sun Z, et al 2012 *Phys Rev Lett* **109**, 016801.

[20]  Liu Z, Liu C X, Wu Y S, Duan W H, Liu F, and Wu J, 2011 *Phys Rev Lett* **107**, 136805.

[21]  Chuang F C, Yao L Z, Huang Z Q, Liu Y T, Hsu C H, Das T, Lin H, and Bansil A, 2014 *Nano Lett* **14**, 2505−2508.

[22]  Ma Y D, Dai Y, Kou L, Frauenheim T, Heine T, 2015 *Nano Lett 15*, 1083.

[23]  Song Z G,  Liu C C,  Yang J B,  Han J Z,  Fu B T,  Ye M,  Yang Y C,  Niu Q,  Lu J, Yao Y G, 2014 *NPG Asia Mater*, **6**, 147.

[24]  Zhou J J, Feng W, Liu C C, Guan S, and Yao Y, 2014 *Nano Lett*, **14**, 4767−4771.

[25]  Drozdov I K, Alexandradinata A, Jeon S, Nadj-Perge S, Ji H, Cava R J, Bernevig B A, and Yazdani A, 2014 *Nat Phys* **10**, 664-669.

[26]  Lu H, and Li S D, 2013 *J Mater Chem C,* **1**, 3677–3680.

[27]  Song Q, Wang B, Deng K, Feng X, Wagner M, Gale J D, Mullen K, and Zhi L, 2013 *J Mater Chem C,* 1, 38-41.

[28]  Liu Y, Wang G, Huang Q, Guo L, and Chen X, 2012 *Phys Rev Lett,* **108**, 225505.

[29]  Wang Q H, Kalantar-Zadeh K, Kis A, Coleman J N, and Strano M S, 2012 *Nat Nanotechnol*, 7, 699–712.

[30]  Nie S M, Song Z, Weng H, and Fang Z, 2015 *Phys Rev B* 91, 235434.

[31]  Sun Y, Felser C, and Yan B, 2015 *arXiv*:1503.08460.

[32] Ma Y, Kou L, Dai Y, and Heine T, 2015 *arXiv*:1504.00197.

[33] Qian X F, Liu J W, Fu L, Li J, 2014 *Science 346*, 1344.

[34] Kresse G, Furthmüller J, 1996 *Phys Rev B 54*, 11169.

[35] Perdew J P, Burke K, Ernzerhof M, 1996 *Phys Rev Lett, 77*, 3865.

[36] Gonze X, Lee C, 1997 *Phys Rev B 55*, 10355.

[37] Togo A, Oba F, Tanaka I, 2008 *Phys Rev B 78*, 134106.

[38] Vogl P, Hjalmarson H P, Dow J D, 1983 *J Phys Chem Solids*, 44, 365-378.

[39] Li L Y, Zhang X M, Chen X, and Zhao M W, 2015 *Nano Lett*, 15, 1296-1301.

[40] Li L Y, and Zhao M W, 2014, *J Phys Chem C*, 118, 19129-19138.

[41] Zhao M W, Chen X, Li L Y, and Zhang X M, 2015 *Sci. Rep.* 5, 8441.

[42] Liu C C, Guan S, Song Z, Yang S A, Yang J, and Yao Y, 2014 *Phys Rev B*, 90, 085431.

[43] Luo W and Xiang H, 2015 *Nano Lett,* 15, 3230−3235.

[44] Chuang F C, Hsu C H, Chen C Y, Huang Z Q, Ozolins V, Lin H, Bansil A, *2013 Appl Phys Lett*, 102, 022424.

[45] Chen X, Li L Y, and Zhang X M, 2015 Phys. Chem. Chem. Phys., 17, 16624.

[46] Zhou M, Ming W, Liu Z, Wang Z, Li P, and Liu F, 2014 *Proc Natl Acad Sci U S A.,* 111, 14378.

[47] Kou L, Chen C F, Smith S, 2015, J Phys. Chem. Lett. 6, 2794.